# Significant reduction of critical currents in MRAM designs using dual free layer with perpendicular and in-plane anisotropy

D. Suess[1], C. Vogler[2], F. Bruckner[1], H. Sepehri-Amin[3], C. Abert[2]

[1] Doppler Laboratory, Institute of Solid State Physics, Vienna University of Technology, 1040 Vienna.

[2] Institute of Solid State Physics, Vienna University of Technology, 1040 Vienna.

[3] National Institute for Materials Science, Tsukuba 305-0047, Japan

*Abstract:* One essential feature in MRAM cells is the spin torque efficiency, which describes the ratio of the critical switching current to the energy barrier. Within this paper it is reported that the spin torque efficiency can be improved by a factor of 3.2 by the use a of dual free layer device, which consists of one layer with perpendicular crystalline anisotropy and a second layer with in-plane crystalline anisotropy. Detailed simulations solving the spin transport equations simultaneously with the micromagnetics equation were performed in order to understand the origin of the switching current reduction by a factor of 4 for the dual layer structure compared to a single layer structure. The main reason could be attributed to an increased spin accumulation within the free layer due to the dynamical tilting of the magnetization within the in-plane region of the dual free layer.

Magnetic random access memory (MRAM) has the potential to become a major player in embedded magnetoresitive non-volatile memory technology due to the fast access and write times. An important design guideline in MRAM development is to reduce the critical switching current while maintaining a high thermal stability. Different designs of spin-transfer torque MRAM (STT-MRAM) cells have been proposed in order to reduce the critical current density. Dual pinned layer were proposed in order to reduce the critical current. Here, the second pinned layer has to point in opposite direction to the first pinned layer and it is separated by a second MgO barrier from the free layer. The critical current was reported to be a factor of 2-3 times smaller than that of a single MgO barrier [1] reaching values of the critical current density of $j_c$ = 0.02 TA/m² as reported in Ref[2]. In this reported structure both pinned layers consist of synthetic antiferromagnetically (SAF) coupled and exchange biased layers leading to layer thicknesses of about 40 nm, which reduces the scalability.

The spin torque efficiency could be improved by thermally assisted switching designs [3]. Here, on top of the free layer an antiferromagnet is deposited which stabilizes the free layer due to exchange bias. If the current is applied, the antiferromagnet is heated above the Néel temperature which decouples the free layer and the antiferromagnet during switching. Hence, the stabilizing support of the antiferromagnet is turned off during switching.

Similar to the dual pinned layer structure an interesting idea is proposed in Ref [4]. Here, the top pinned layer is designed to be fixed in a direction perpendicular to the bottom pinned layer, which has perpendicular anisotropy. Micromagnetic simulations predict that due to the precessional switching, current pulses as short as 50 ps can be used in order to switch the structure. The critical current for the proposed structure was about 2mA which corresponds to a critical current density of about 1 TA/m² for the given lateral dimension. In these simulations the magnetization of the pinned layer was assumed to be fixed in one direction.

Various attempts have been made in order to employ an exchange-spring concept similar to hard disk media [5,6] in order to reduce the switching current and maintain high thermal stability. However, the possible reduction in critical current density with the help of exchange spring effect was reported to be only 20% at the cost of longer switching times [7]. Recently, more work was presented on the switching of exchange-spring like free layers using the Slonczewski spin-torque term. However, if an inhomogeneous state between the soft and hard part of the free layer is formed, the applicability of this simple approach is questionable and the studies presented significantly different reduction of critical currents. For example, in the simple Slonczewski formalism, the effective spin-transfer torque due to the dual free layer is hard to model since it is not clear to which degree the spins become polarized by the first free layer resulting in an uncertain additional torque on the second free layer.

In this letter we will report a drastic switching current reduction by using one free layer with strong perpendicular anisotropy coupled to a second free layer with moderate in-plane anisotropy that allows to rotate the magnetization during the switching process. This structure is beneficial for a significant switching current reduction and easier to realize than the structure proposed in Ref [4] since not two easy axis with 90° angle between them is required. The in-plane structure can be realized in a most simple structure just by the shape anisotropy of a soft magnet with zero crystalline anisotropy.

The used model is based on the spin drift diffusion model as described by Zhang et al. [8] . The coupling of magnetization dynamics and solution of spin drift-diffusion equation according to Abert et al. [9] is used within the paper. Here, besides the magnetization as function of time, the spin accumulation *s* and the electrical potential *u* are calculated as function of input currents. The used model is a continuous model, where the magnetization as well as the spin accumulation are continuous function of space and all investigated layers are discretized using a finite element method [10,11].

As shown in *Fig. 1*, the structure consists of: lead$_1$ = 10 nm, fixed = 5 nm, non-magnetic layer NML = 1 nm free$_1$ = 2nm, exchange break layer EBL = 1.0 nm, free$_2$ = 1.5 nm and lead$_2$ = 10 nm. The diameter is *d* = 20 nm.

The magnetic and electric material parameters of the fixed layer are: anisotropy constant $K_1$ = 10 MJ/m³, saturation polarization $J_s$ = 1.0 T, exchange constant A = 10 pJ/m, damping constant $\alpha$ = 0.02 and the exchange strength between the conducting electrons and magnetization J = 4.1x10$^{-20}$ J, dimensionless polarization parameters $\beta$ = 0.9, $\beta'$ = 0.8, the spin flip relaxation time $\tau_{sf}$ = 5x10$^{-14}$ s, the diffusion constant $D_0$ = 10$^{-3}$ m²/s . The easy axis is aligned 5,7° off the *z*-axis in order to break the symmetry and to avoid perfect antiparallel orientation between the magnetization in the free layer and the fixed layer. The large value of the anisotropy constant was used to pin the fixed layer in one direction. It might be important to note that a decrease of the anisotropy constant to $K_1$ = 3 MJ/m³ does not change the results.

The parameters for the free layers are the same as for the fixed layer, except for the anisotropy constant and coupling strength between free$_1$ and free$_2$. The anisotropy of the layer free1 is in all simulations $K_1$ = 0.4 MJ/m³ and the easy axis *k* = (0,0,1). In this paper, the fundamental features of the described effect are studied. It does not focus on one particular material, and thus for simplicity the electronic parameters of the non-magnetic layer (NML), exchange breaking layer (EBL) and leads are assumed to be the same as for the pinned layer. Since, the used model uses the spin-diffusion model, it is strictly valid only for metallic structures, using conducting layers. Interesting structures which could be directly modeled are for example Heusler alloys [12] which show a strong degree of spin polarization and could be a potential candidate for MRAM cells. Higher currents can be applied at which the tunnel barrier degeneration for large currents has not to be considered. The physics of the coupled free layers is well described by the used spin-diffusion model and goes beyond simple Slonczewski terms as described in Ref [9,13]. If tunnel barriers are important, the physics of the barrier can be approximated by appropriate effective material parameters within the NML. The diffusive processes in the metallic parts of the structure remains unchanged and important to be resolved [14].

In *Fig. 1* an electric current with a strength of j = 0.3 TA/m² is applied along the *z*-direction. The positive current represents electrons flowing from lead$_2$ towards lead$_1$. Initially the magnetization of fixed and free$_1$ layer is pointing in the +z direction. In the simulation, the magnetization in the layer free$_2$ is fixed with a strong anisotropy of $K_1$ = 0.6 MJ/m³ along the *k* = (0,1,0) direction. Due to the parallel magnetization between fixed and free$_1$ layer, no particular accumulation of spins (spin accumulation) is observed at this interface as shown in *Fig. 1*. The reason is that in the spin drift-diffusion model, spin accumulation arises due to gradients of the magnetization. In contrast, strong spin accumulation is observed at the free$_1$/free$_2$ interface as well as at the free$_2$/lead$_2$ interface.

The effect of the strength of the anisotropy in free$_2$ on the switching current is shown in *Fig. 2* (a). Here, a current density with a rise time of $r$ = 1 TA/(m²ns) is applied after the system is equilibrated at zero current. In all simulation presented in Fig. 2, the exchange coupling in the 1 nm thick EBL between free$_1$ and free$_2$ is $A$ = 0.1 pJ/m. In the case of positive values of $K_1$, the easy axis of the layer free$_2$ is the *y*-axis $k$ = (0,1,0). For comparison, a simulation without the layer free$_2$ and the EBL is shown in *Fig. 2* (a). In the case of $K_1$ = 0.6 MJ/m³ , the magnetization in the layer free$_2$ is strongly aligned with the *y*-direction. For this design the critical current can be reduced to $j$ = 0.93 TA/m² compared to $j$ = 1.27 TA/m² for the single free layer. This structure is similar to the structure reported in Ref [4].   It is very interesting to note that a further reduction of about a factor of 1.6 can be realized by reducing the anisotropy constant of the layer free$_2$ to $K_1$ = 0.4 MJ/m³.  From a practical point of view, it is not trivial to realize an MRAM like structure where the magnetic layer free$_1$ has strong perpendicular anisotropy and free$_2$, which is separated by only a thin EBL, shows an uniaxial anisotropy in a perpendicular direction.

For comparison two simulations are performed where the damping constant is reduced to $\alpha$ = 0.005 in all magnetic layer. If the damping constant is reduced in the structure consisting only of one free layer (Fig. 2 left) the critical current decreases by about 7%. For the structre of the composite free layer with one layer with inplane anistropy (Fig. 2 right) the damping constant almost does not affect the critical switching current.

Within this paper we study the possibility to realize a highly functional structure with a layer free$_2$ having an in-plane anisotropy. In the simplest design, a soft magnetic layer free$_2$ with zero crystalline anisotropy, $K_1$ = 0.0 MJ/m³ leads to an in-plane magnetization due to the shape anisotropy. In *Fig. 2* (b) the switching process for different intrinsic in-plane anisotropy constants is shown. The in-plane anisotropy is realized by a negative $K_1$ value and an easy axis pointing in the $k$ = (0,0,1) direction. It can be seen that with increasing magnitude of the in-plane anisotropy the critical current can be decreased. For comparison also data for positive  $K_1$ values and  $k$ = (0,0,1) are shown, which represent an exchange spring like structure of a hard magnetic layer free$_1$ and a softer layer free$_2$ similar to the structures investigated in Ref [7]. It can be seen that the in-plane anisotropy decreases the switching current more efficiently than the soft magnetic layer free$_2$. For comparison the case of the in-plane anisotropy $K_1$ = -0.3 MJ/m³ is also plotted in *Fig. 2* (a).

In order to find the origin of the switching current reduction, if an in-plane anisotropy is used (*Fig. 3* bottom) instead of a strong uniaxial anisotropy in the *k* = (0,1,0) direction (*Fig. 3* top), the transient states during reversal are investigated. It can be seen that for the case of the moderate in-plane anisotropy, the magnetization in the layer free$_2$ starts to align into the positive *z*-direction, if a positive current is applied. This effect can be easily understood by conventional Slonczewski spin-transfer torque. Due to the orientation of the layer free$_2$, which rotates into the antiparallel direction with respect to layer free$_1$, the spin accumulation increases at the interface between free$_1$ and free$_2$ as shown in *Fig. 1* (red dotted line). In a simple picture, the spin accumulation can be regarded as an additional contribution to the effective field. Hence, the strong negative *z*-component of the spin accumulation leads to switching of the layer free$_1$ into the –z direction.

The effect of reducing the critical current by this flexible second free layer also works if the current direction is reversed. Here, the layer free$_2$ will be rotated into the parallel alignment with respect to the layer free$_1$. The back scattered electrons then support the switching of the layer free$_1$ as it can be also understood in the Slonczewski picture.

The critical current can also be reduced if the uniaxial anisotropy which points into the *k* = (0,1,0) direction is reduced to moderate values as it can be seen in *Fig. 2* (a). One effect which seems to depend on the dynamics of the system which requires further consideration is, why the critical current remains small even for large negative in-plane anisotropy. Transient states during switching indicate that - as expected - the magnetization in the layer free$_2$ is kept indeed mostly in-plane. However, in contrast to the case of the strong uniaxial anisotropy in k = (0,1,0), the magnetization in layer free$_2$ starts to rotate within the easy plane, which seems to support the switching of the layer free$_1$.

In order to investigate the optimum coupling strength between the two free layers simulations are performed where the coupling is varied in the 1 nm thick EBL. It is interesting to note that there is no optimum coupling strength but a sufficient low coupling leads to the smallest critical currents as shown in *Fig. 4*.

In all previous simulations the current was applied with a rise time of *r* = 1 TA/(m²ns). Since, in MRAM devices the switching speed is a critical issue, we present simulations where various current pulses are applied in the following. We apply a current pulse which increases linearly with *r* = 1 TA/(m²ns) until the maximum current *j* is reached. This current *j* is kept constant for a time τ. After this time, the current is decreased with *r* = 1 TA/(m²ns). In *Fig. 5* the final state of the magnetization in the layer free$_1$ after this current pulse is color coded. Red represents the state pointing in +z direction and blue in –z direction. In the top and bottom pictures the initial state in the layer free1 was pointing in the +z and –z direction, respectively. An impressive reduction of the critical switching current from about $j$ = 0.8 TA/m² (single free layer) to $j$ = 0.2 TA/m² (dual free layer) can be observed for example for $\tau$ = 0.3 ns.

For the switching from down to up, the single free layer and the dual free layer show similar performance. In this case for both designs a strong spin accumulation arises at the interface fixed layer/free$_1$ layer, which dominates the reversal process.

Finally, the thermal stability of the dual free layer structure is compared with the single free layer design in order to be able to calculate the spin torque efficiency $\eta = \Delta E/I_c$. The energy barrier is calculated using the string method[15] which showed similar performance as the nudged elastic band method [16,17], but is easier to implement. In *Fig. 6* the most probable transition path is compared for the single free layer (black line) and the dual free layer (dotted red line). The energy barrier of the single free layer and dual free layer is *ΔE = 18.8 k$_B$T$_{300}$* and *ΔE = 15.1 k$_B$T$_{300}$*, respectively.

To conclude, it is shown that the critical switching currents can be significantly reduced in MRAM devices if adjacent to the free layer a second free layer with in-plane anisotropy is deposited. The critical switching current for switching from magnetization up to down of the free layer can be reduced by a factor of 4. The energy barrier only slightly changes, which leads to an improved spin-torque efficiency $\eta = \Delta E/I_c$ of the dual layer design by a factor of 3.2.

The financial support by the Austrian Federal Ministry of Science, Research and Economy and the National Foundation for Research, Technology and Development as well as the Austrian Science Fund (FWF) under Grant Nos. F4112 SFB ViCoM, I2214-N20 and the Vienna Science and Technology Fund (WWTF) under Grant No. MA14-044 is acknowledged.

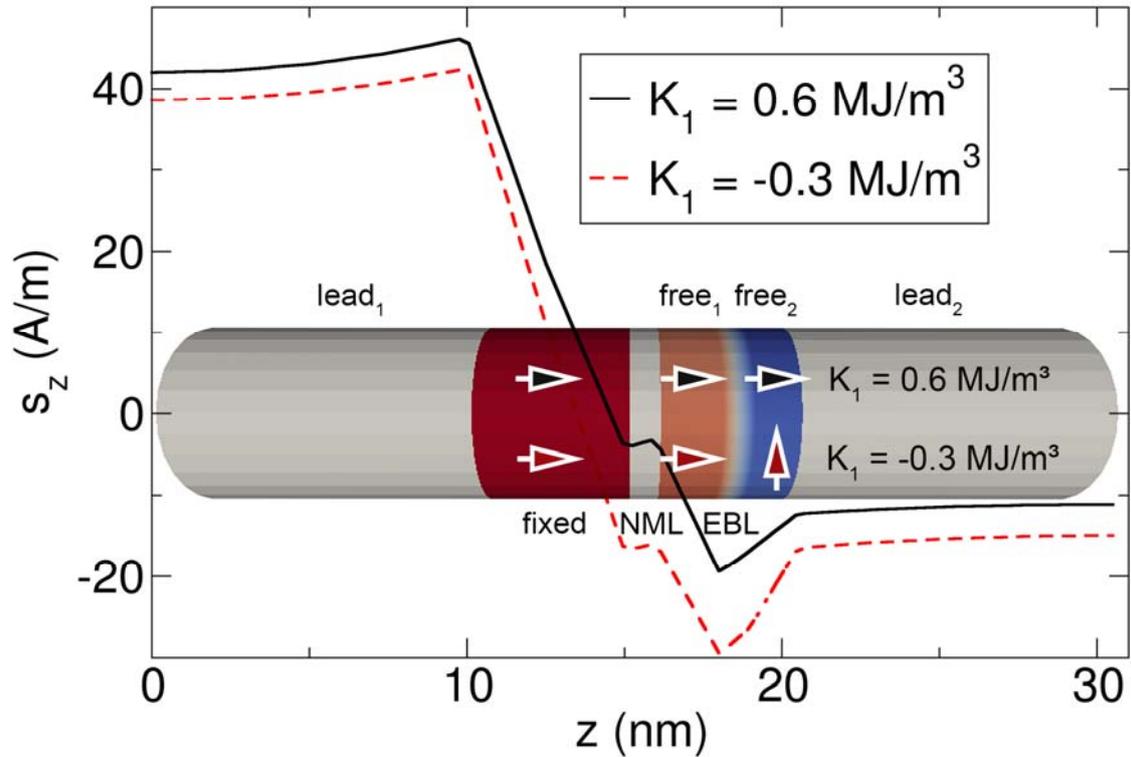

Fig. 1: The *z*-component of the spin accumulation is plotted. The layers lead$_1$, non-magnetic layer (NML) and lead$_2$ are non-magnetic layers. The layers fixed, free$_1$ and free$_2$ are magnetic layers. (black solid line): The magnetization in the free$_2$ layer is fixed with a strong uniaxial anisotropy of $K_1$ = 0.6 MJ/m³ in the k=(0,1,0) direction. (red dotted): An in-plane anisotropy is realized in the free$_2$ layer with $K_1$ = -0.3 MJ/m³ and the easy axis is k=(0,0,1). The current is j = 0.3 TA/m² . The equilibrium magnetization is sketched in the magnetic layer for the two cases with the arrows.

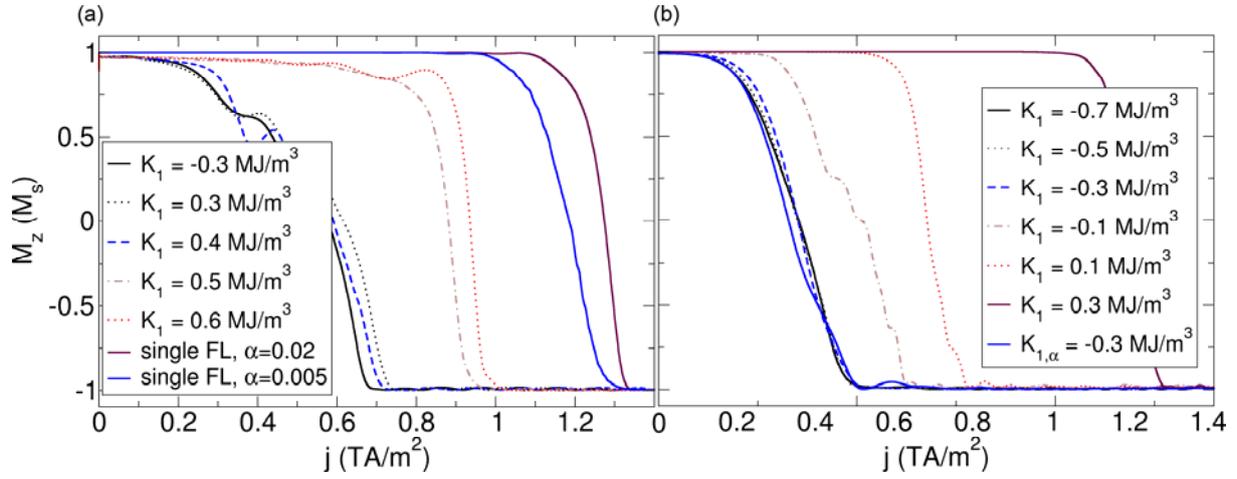

Fig. 2: The *z*-component of the magnetization in the layer free$_1$ is plotted during switching from up to down as function of applied electric current. Current rise time is r = 1TA/(m²ns). The exchange constant in the EBL is *A* = 0.1 pJ/m. In the simulation single FL, $\alpha$ = 0.005 (left) and $K_{1,\alpha}$ = -0.3 MJ/m³ (right) the damping constant is set to $\alpha$ = 0.005 in all magnetic layers. In all other simulations in the paper $\alpha$ = 0.02 in all magnetic layers. (a) For simulation with $K_1$ = -0.3 MJ/m³ the easy axis is pointing in *k* = (0,0,1). For all other simulations with positive anisotropy constant *k* = (0,1,0). Hence, in all cases the magnetization in the very top layer is parallel to the plane. (single free layer) For comparison the switching process for a free layer is shown that consists only of the layer free$_1$ and does not have the layer free$_2$. (b) The anisotropy constant in the layer free$_2$ is varied. For all simulations the anisotropy axis is *k* = (0,0,1).

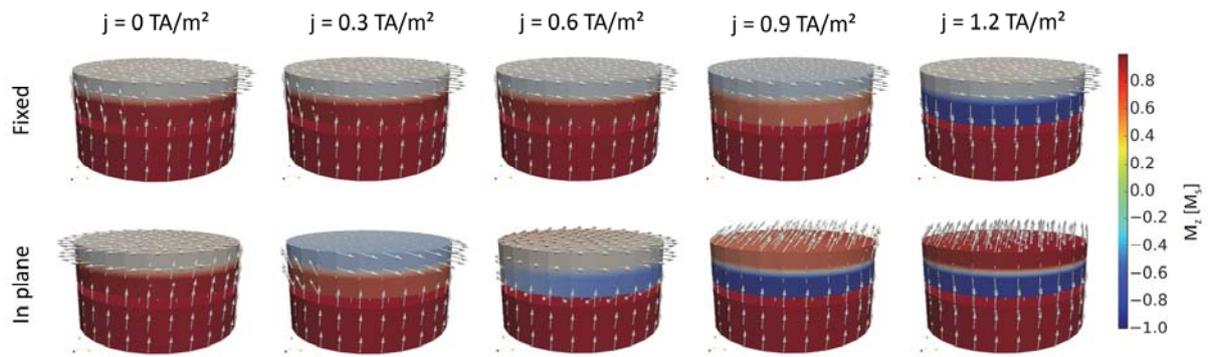

Fig. 3: Transient states during switching for (top) a dual free layer, where the very top layer (free$_2$) has a fixed uniaxial anisotropy with $K_1$ = 0.6 MJ/m³ and $k$ = (0,1,0) . The $z$-component of the magnetization is color coded. (bottom) a dual free layer, where free$_2$ has an in-plane anisotropy with $K_1$ = -0.3 MJ/m³ and $k$ = (0,0,1). The exchange constant in the EBL is $A$ = 0.1 pJ/m.

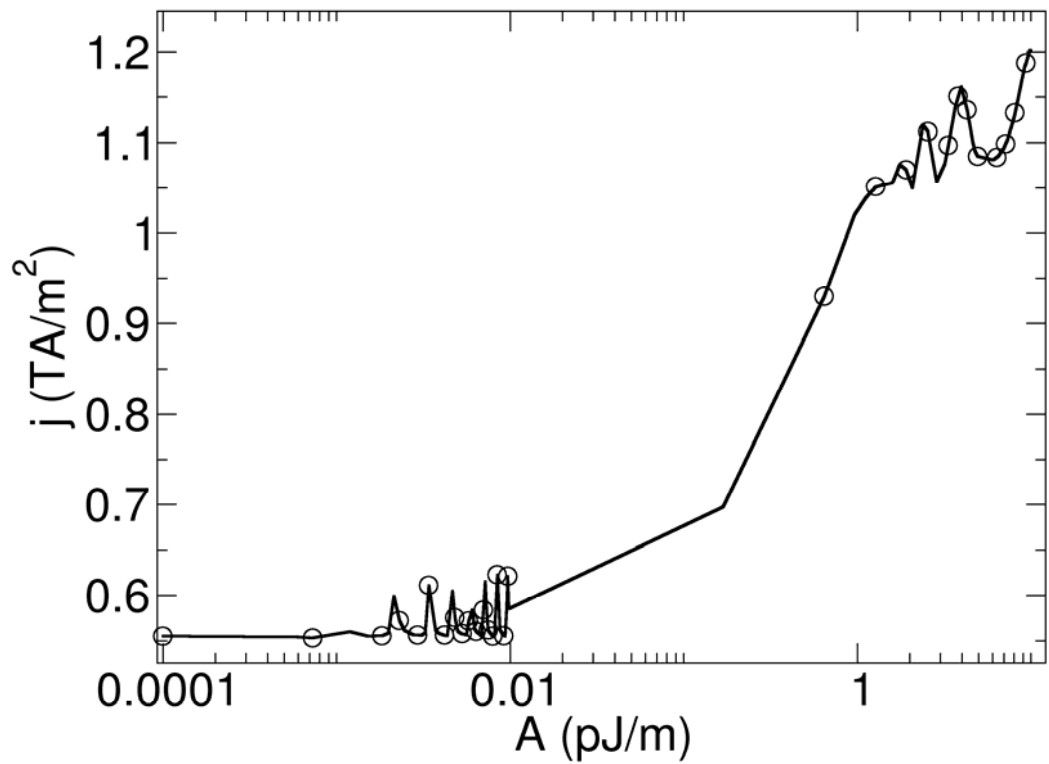

Fig. 4: Dependence of switching current from up to down as function of the exchange coupling between the two free layers. The critical current was determined from a current hysteresis loop with *r* = 1 TA/(m²ns)

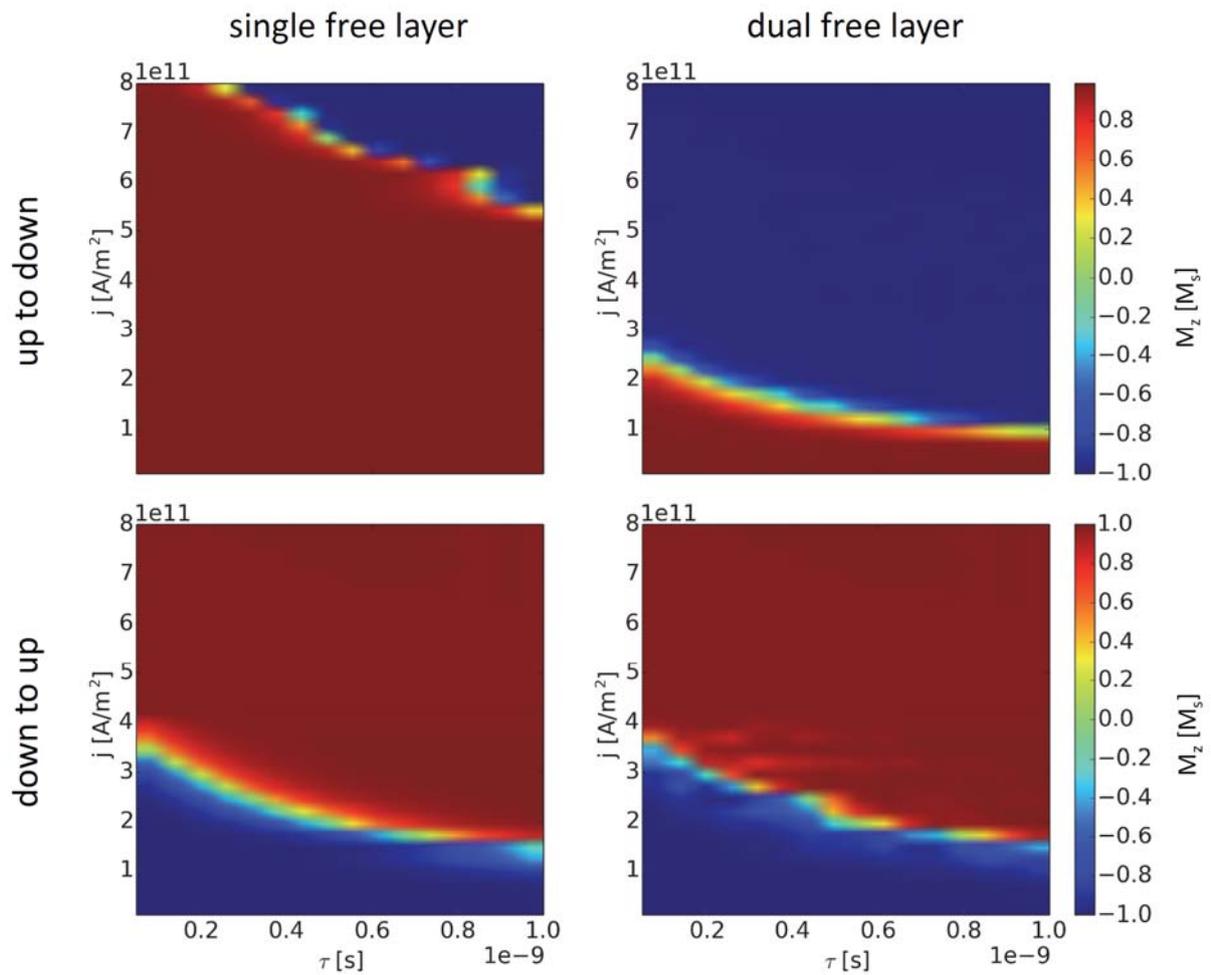

Fig. 5: Phase plots where the color code shows the z-component of the magnetization in the layer $free_1$ after a current pulse with pulse length $\tau$ and current strength $j$ is applied. (left column) single free layer (right column) dual free layer design with $K_1$ = -0.3 MJ/m³ in $free_2$. The exchange constant in the EBL is $A$ = 0.1 pJ/m

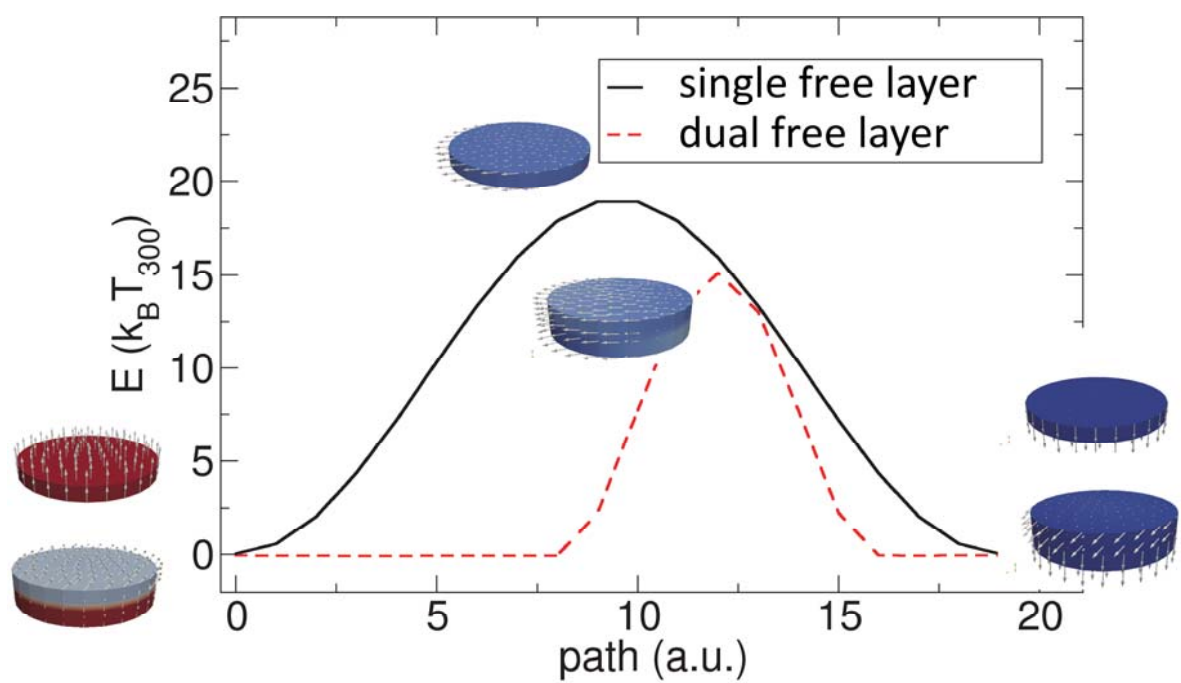

Fig. 6: (red dotted) Energy barrier simulation ($\Delta E = 15.1\ k_BT_{300}$) of the dual free layer structure with K1 = -0.3 MJ/m³ and A = 0.01 pJ/m³. (black line) Energy barrier simulation for the device that consists only of layer free$_1$ ($\Delta E = 18.8\ k_BT_{300}$).